\documentclass[aps,tightenlines,prl,twocolumn,superscriptaddress,showpacs,amsmath,amssymb,preprintnumbers,footinbib,floatfix]{revtex4}

\usepackage{graphicx}
\usepackage{dcolumn}
\usepackage{array}

\newcommand{\mbc}{{M_{\textrm{bc}}}}
\newcommand{\deltae}{{\Delta E}}
\newcommand{\BR}{{\mathcal B}}
\newcommand{\ebeam}{{{E_{\textrm{b}}^{\ast}}}}
\newcommand{\lint}{{L_{\textrm{int}}}}
\newcommand{\mev}{{\hbox{ MeV}}}
\newcommand{\gev}{{\hbox{ GeV}}}
\newcommand{\mmev}{{\hbox{ MeV}/c^2}}
\newcommand{\mgev}{{\hbox{ GeV}/c^2}}
\newcommand{\bs}{{B_s^0}}
\newcommand{\bsst}{{B_s^{\ast}}}
\newcommand{\barbsst}{{{\bar B_s}^{\ast}}}
\newcommand{\bsST}{{B_s^{(\ast)}}}
\newcommand{\bsSTbsST}{{\bsST{\bar B_s}^{(\ast)}}}
\newcommand{\KS}{{K_S^0}}
\newcommand{\ds}{{D_s^-}}
\newcommand{\bsdspi}{{\bs\to D_s^-\pi^+}}
\newcommand{\bsdsstpi}{{\bs\to D_s^{\ast-}\pi^+}}
\newcommand{\bsdsSTpi}{{\bs\to D_s^{(\ast)-}\pi^+}}
\newcommand{\bsdsk}{{\bs\to D_s^{\mp}K^{\pm}}}
\newcommand{\thetahel}{\theta_{\textrm{hel}}}
\newcommand{\bfbstodspiun}{[3.67^{+0.35}_{-0.33}({\textrm{stat.}}){}^{+0.43}_{-0.42}({\textrm{syst.}})}
\newcommand{\bfbstodspideux}{\pm0.49(f_s)]\!\times\!10^{-3}}
\newcommand{\bfbstodspi}{\bfbstodspiun\bfbstodspideux}
\newcommand{\bfbstodskun}{[2.4^{+1.2}_{-1.0}({\textrm{stat.}})}
\newcommand{\bfbstodskdeux}{{\pm0.3}({\textrm{syst.}})\pm0.3(f_s)]\!\times\!10^{-4}}
\newcommand{\bfbstodsk}{\bfbstodskun\bfbstodskdeux}

\newcommand{\ratio}{{\left(6.5^{+3.5}_{-2.9}\right)\%}}

\newcommand{\fss}{{\left(90.1^{+3.8}_{-4.0}\pm0.2\right)\%}}
\newcommand{\fs}{{\left(7.3^{+3.3}_{-3.0}\pm0.1\right)\%}}
\newcommand{\f}{{\left(2.6^{+2.6}_{-2.5}\right)\%}}

\newcommand{\mbs}{{\left(5364.4\pm1.3\pm0.7\right)\mmev}}
\newcommand{\mbsst}{{\left(5416.4\pm0.4\pm0.5\right)\mmev}}
\newcommand{\deltambsstmbs}{{52.0\pm1.5\mmev}}

\newcommand{\yieldbsstbssttodspi}{{145^{+14}_{-13}}}
\newcommand{\dspisignbsstbsst}{{21.0\,\sigma}}

\newcommand{\yieldbsstbstodspi}{{11.8^{+5.8}_{-5.0}}}
\newcommand{\dspisignbsstbs}{{2.7\,\sigma}}

\newcommand{\yieldbsbstodspi}{{4.0^{+4.6}_{-3.7}}}
\newcommand{\dspisignbsbs}{{1.1\,\sigma}}

\newcommand{\yieldbstodsk}{{6.7^{+3.4}_{-2.7}}}
\newcommand{\dsksign}{{3.5\,\sigma}}
\newcommand{\etal}{{\textit{et al.}}}
\begin{document}
  
     \preprint{\vbox{
         \hbox{KEK Preprint 2008-27}
        \hbox{BELLE Preprint 2008-24}
       }
     }
  
  \title{Measurement of the Decay \boldmath{$\bsdspi$} and Evidence for \boldmath{$\bsdsk$}
    in \boldmath{$e^+e^-$} Annihilation at \boldmath{$\sqrt s\sim10.87\gev$}}
  
  \affiliation{Budker Institute of Nuclear Physics, Novosibirsk}
  \affiliation{Chiba University, Chiba}
  \affiliation{University of Cincinnati, Cincinnati, Ohio 45221}
  \affiliation{T. Ko\'{s}ciuszko Cracow University of Technology, Krakow}
  \affiliation{Justus-Liebig-Universit\"at Gie\ss{}en, Gie\ss{}en}
  \affiliation{The Graduate University for Advanced Studies, Hayama}
  \affiliation{Hanyang University, Seoul}
  \affiliation{University of Hawaii, Honolulu, Hawaii 96822}
  \affiliation{High Energy Accelerator Research Organization (KEK), Tsukuba}
  \affiliation{Institute of High Energy Physics, Chinese Academy of Sciences, Beijing}
  \affiliation{Institute of High Energy Physics, Vienna}
  \affiliation{Institute of High Energy Physics, Protvino}
  \affiliation{Institute for Theoretical and Experimental Physics, Moscow}
  \affiliation{J. Stefan Institute, Ljubljana}
  \affiliation{Kanagawa University, Yokohama}
  \affiliation{Korea University, Seoul}
  \affiliation{Kyungpook National University, Taegu}
  \affiliation{\'Ecole Polytechnique F\'ed\'erale de Lausanne (EPFL), Lausanne}
  \affiliation{Faculty of Mathematics and Physics, University of Ljubljana, Ljubljana}
  \affiliation{University of Maribor, Maribor}
  \affiliation{University of Melbourne, School of Physics, Victoria 3010}
  \affiliation{Nagoya University, Nagoya}
  \affiliation{Nara Women's University, Nara}
  \affiliation{National Central University, Chung-li}
  \affiliation{National United University, Miao Li}
  \affiliation{Department of Physics, National Taiwan University, Taipei}
  \affiliation{H. Niewodniczanski Institute of Nuclear Physics, Krakow}
  \affiliation{Nippon Dental University, Niigata}
  \affiliation{Niigata University, Niigata}
  \affiliation{University of Nova Gorica, Nova Gorica}
  \affiliation{Osaka City University, Osaka}
  \affiliation{Osaka University, Osaka}
  \affiliation{Panjab University, Chandigarh}
  \affiliation{Saga University, Saga}
  \affiliation{University of Science and Technology of China, Hefei}
  \affiliation{Seoul National University, Seoul}
  \affiliation{Sungkyunkwan University, Suwon}
  \affiliation{University of Sydney, Sydney, New South Wales}
  \affiliation{Tata Institute of Fundamental Research, Mumbai}
  \affiliation{Toho University, Funabashi}
  \affiliation{Tohoku Gakuin University, Tagajo}
  \affiliation{Department of Physics, University of Tokyo, Tokyo}
  \affiliation{Tokyo Institute of Technology, Tokyo}
  \affiliation{Tokyo Metropolitan University, Tokyo}
  \affiliation{Tokyo University of Agriculture and Technology, Tokyo}
  \affiliation{Virginia Polytechnic Institute and State University, Blacksburg, Virginia 24061}
  \affiliation{Yonsei University, Seoul}
  
  \author{R.~Louvot}\affiliation{\'Ecole Polytechnique F\'ed\'erale de Lausanne (EPFL), Lausanne}
  \author{J.~Wicht}\affiliation{High Energy Accelerator Research Organization (KEK), Tsukuba}
  \author{O.~Schneider}\affiliation{\'Ecole Polytechnique F\'ed\'erale de Lausanne (EPFL), Lausanne}
  
  \author{I.~Adachi}\affiliation{High Energy Accelerator Research Organization (KEK), Tsukuba} 
  \author{H.~Aihara}\affiliation{Department of Physics, University of Tokyo, Tokyo} 
  \author{K.~Arinstein}\affiliation{Budker Institute of Nuclear Physics, Novosibirsk} 
  \author{V.~Aulchenko}\affiliation{Budker Institute of Nuclear Physics, Novosibirsk} 
  \author{T.~Aushev}\affiliation{\'Ecole Polytechnique F\'ed\'erale de Lausanne (EPFL), Lausanne}\affiliation{Institute for Theoretical and Experimental Physics, Moscow} 
  \author{A.~M.~Bakich}\affiliation{University of Sydney, Sydney, New South Wales} 
  \author{V.~Balagura}\affiliation{Institute for Theoretical and Experimental Physics, Moscow} 
% ...

  \collaboration{The Belle Collaboration}
  \noaffiliation
  
  \begin{abstract}
    
    We have studied $\bsdspi$ and $\bsdsk$ decays  using 23.6 fb$^{-1}$ of data collected  at the $\Upsilon(5S)$ resonance with
    the Belle detector at the KEKB $e^+e^-$ collider.
    This highly pure $\bsdspi$ sample is used to measure the branching fraction, $\BR(\bsdspi)=$ $\bfbstodspi$ ($f_s=N_{\bsSTbsST}/N_{b\bar b}$)
    and the fractions of $\bs$ event types at the $\Upsilon(5S)$ energy, {in particular $N_{\bsst\barbsst}/N_{\bsSTbsST}=\fss$. 
    We also determine the masses $M(\bs)=\mbs$ and $M(\bsst)=\mbsst$.}
    {In addition,} we observe $\bsdsk$ decays with a significance of $\dsksign$ and measure  ${\cal B}(\bsdsk)=\bfbstodsk$.
    
  \end{abstract}
  
  
  \maketitle
      {
	\renewcommand{\thefootnote}{\fnsymbol{footnote}}}
      \setcounter{footnote}{0}
      
      The decay $\bsdspi$ \cite{chargeconj} has a relatively large branching fraction and
      is a primary normalization mode at hadron colliders,
      where the absolute production rate of $\bs$ mesons is difficult to measure directly.
      It proceeds dominantly via a  Cabibbo-favoured tree process.
      The decay $B^0\to D^-\pi^+$ proceeds through the same tree process but may also have additional contributions from $W$-exchange,
      so a comparison of the partial widths of the two decays can give insight into the poorly known $W$-exchange process.
      The Cabibbo-suppressed mode $\bsdsk$ is mediated by $b\to c$ and $b\to u$ tree transitions of similar order
      ($\sim\lambda^3$, in the Wolfenstein parameterization \cite{PRL_51_1945}),
      which raises the possibility of measuring time-dependent $CP$-violating effects \cite{ZPHYSC_54_653}.
      It has recently become possible to produce $\bs$ events from $e^+e^-$ collisions at the $\Upsilon$(5S) resonance
      in sufficiently large numbers to achieve interesting and competitive measurements.
      $\Upsilon$(5S) events may also be used to  determine precisely the masses of  $\bsst$ and $\bs$;
      the mass difference can be compared with that of $B^{\ast0}$ and $B^0$ to test heavy-quark symmetry
      \cite{PRD_68_054024}, which predicts equality between them.
      Properties of the $\Upsilon(5S)$ such as the fraction of events containing a $\bs$ and
      the relative proportions of $\bs\bar\bs$, $\bsst\bar\bs$, and $\bsst\barbsst$ provide additional tests of
      heavy quark theories \cite{PRD_55_272,PRL_53_878}.
      
      In this Letter, we report measurements performed with fully reconstructed $\bsdspi$ and $\bsdsk$ decays
      in $\lint=\left(23.6\pm0.3\right)$~fb$^{-1}$ of data collected with the Belle detector at
      the KEKB asymmetric-energy (3.6\gev~on 8.2\gev) $e^+e^-$ collider \cite{kekb} operated at the $\Upsilon(5S)$ resonance.
      {
	The beam energy in the center-of-mass (CM) frame is measured to be $\ebeam=\sqrt{s}/2=5433.5\pm0.5\mev$ with $\Upsilon(5S)\to\Upsilon(1S)\,\pi^+\,\pi^-$, $\Upsilon(1S)\to\mu^+\mu^-$ decays \cite{PRL_100_112001}}.
      The total $b\bar b$ cross section at the $\Upsilon(5S)$ energy has been measured to be
      $\sigma_{b\bar b}^{\Upsilon(5S)}=\left(0.302\pm0.014\right)$~nb \cite{PRL_98_052001},
      which includes $B^0$, $B^+$ and $\bs$ events.
      Three $\bs$ production modes are kinematically allowed: $\bs\bar\bs$, $\bsst\bar\bs$, and $\bsst\bar\bsst$.
      The $\bsst$ decays electromagnetically to $\bs$, emitting a photon with  energy  $E_{\gamma}\sim53\mev$.
      The fraction of $b\bar b$ events containing a $\bsSTbsST$ pair has been measured to be
      $f_s=N_{\bsSTbsST}/N_{b\bar b}=(19.5^{+3.0}_{-2.3})\%$ \cite{PRL_98_052001}.
      The number of $\bs$ mesons in the sample is thus
      $N_{\bs}=2\times\lint\times\sigma_{b\bar b}^{\Upsilon(5S)}\times f_s=(2.78^{+0.45}_{-0.36})\!\times\!10^6$.
      The $\bs$ production mode ratios are defined as
      $f_{\bsst\barbsst}=N_{\bsst\barbsst}/$ $N_{\bsSTbsST}$, $f_{\bsst\bar\bs}=N_{\bsst\bar\bs}/N_{\bsSTbsST}$
      and $f_{\bs\bar\bs}=N_{\bs\bar\bs}/N_{\bsSTbsST}$.
      Belle previously measured $f_{\bsst\barbsst}=\left(93^{+7}_{-9}\right)\%$ \cite{PRD_76_012002}.
   
      The Belle detector is a large-solid-angle magnetic
      spectrometer that consists of a silicon vertex detector,
      a central drift chamber (CDC), an array of
      aerogel threshold Cherenkov counters (ACC),
      a barrel-like arrangement of time-of-flight
      scintillation counters (TOF), and an electromagnetic calorimeter
      comprised of CsI(Tl) crystals located inside
      a superconducting solenoid coil that provides a 1.5~T
      magnetic field. An iron flux-return located outside of 
      the coil is instrumented to detect $K_L^0$ and to identify 
      muons. The detector 
      is described in detail elsewhere~\cite{beld}.

      Reconstructed charged tracks are required to have a maximum impact parameter with respect
      to the nominal interaction point of 0.5~cm in the radial direction and 3~cm in the beam-axis direction.
      A likelihood ratio $\mathcal R_{K/\pi}=\mathcal{L}_K/\left(\mathcal L_{\pi}+\mathcal L_K\right)$ is built using
      ACC, TOF and CDC ($dE/dx$) measurements.
      A track is identified as a pion if $\mathcal R_{K/\pi}<0.6$ or as a kaon otherwise.
      With this selection, the identification efficiency for pions (kaons) is about $91\%$ ($85\%$), while the fake rate is about $9\%$ ($14\%$).
      
      Neutral kaons are reconstructed via the decay $\KS\to\pi^+\pi^-$ with no identification requirements for the two charged pions.
      The $\KS$ candidates are required to have an invariant mass within $\pm7.5\mmev$ ($\pm4\,\sigma$) of
      the nominal $\KS$ mass (all nominal mass values are taken from Ref.~\cite{pdg}).
      Requirements on the $\KS$ vertex displacement from the interaction point and on the difference between vertex and $\KS$ flight directions are applied.
      The criteria are described in detail elsewhere \cite{phd_ffang}.
      The $K^{\ast0}$ ($\phi$) candidates are reconstructed via the decay $K^{\ast0}\to K^+\pi^-$ ($\phi\to K^+K^-$) with an invariant mass
      within $\pm50\mmev$ ($\pm12\mmev$) of the nominal 
      mass.
      
      Candidates for $\ds$ are reconstructed in the three modes $\ds\to\phi\pi^-$, $\ds\to K^{\ast0}K^-$, and $\ds\to\KS K^-$ and
      required to have mass within $\pm15\mmev$ ($\pm3\,\sigma$) of the nominal $\ds$  mass for $\bsdspi$ and within $\pm8\mmev$ for $\bsdsk$.
      {Following Ref.~\cite{PRD_76_012002},} the signals for $\bsdspi$ and $\bsdsk$ are observed using two variables: the beam-constrained mass of the $\bs$ candidate
      $\mbc=\sqrt{\ebeam^2-\vec p_{\bs}^{\ast2}}$ and the energy difference $\deltae=E_{\bs}^{\ast}-\ebeam$,
      where $(E_{\bs}^{\ast},\vec p_{\bs}^{\ast})$ is the four-momentum of the $\bs$ candidate
      expressed in the CM frame.
      We select candidates with $\mbc>5.3\mgev$ and $-0.3\gev<\deltae<0.4\gev$.
      In each event the $\bs$ candidate with the $\ds$ mass closest to its nominal value is selected for further analysis; 
      only $\approx1\%$ of events have more than one candidate.
      
      Further selection criteria are developed using Monte Carlo (MC) samples based on EvtGen \cite{evtgen} and GEANT \cite{geant} detector simulation.
      The most significant source of background is continuum events, $e^+e^-\to u\bar u, d\bar d, s\bar s, c\bar c$.
      In addition, for the $\bsdsk$ mode there is also a large background from $\bsdspi$, where the $\pi^+$ is misidentified as a $K^+$.
      The expected continuum background, $N_{\textrm{bkg}}$, is estimated using MC-generated continuum events representing three times the data.
      The expected signal, $N_{\textrm{sig}}$, is obtained assuming $\BR\left(\bsdspi\right)=3.0\!\times\!10^{-3}$ and
      $f_{\bsst\barbsst}=93\%$ for the $\bsdspi$ analysis and
      $\BR\left(\bsdsk\right)=3.7\!\times\!10^{-4}$  for  the $\bsdsk$ analysis.
      For $\bsdsk$, we {assume the values of $\BR(\bsdspi)$ and $f_{\bsst\barbsst}$ obtained in the $\bsdspi$ analysis}.
      
\begin{figure}[t]
	\includegraphics[width=\linewidth]{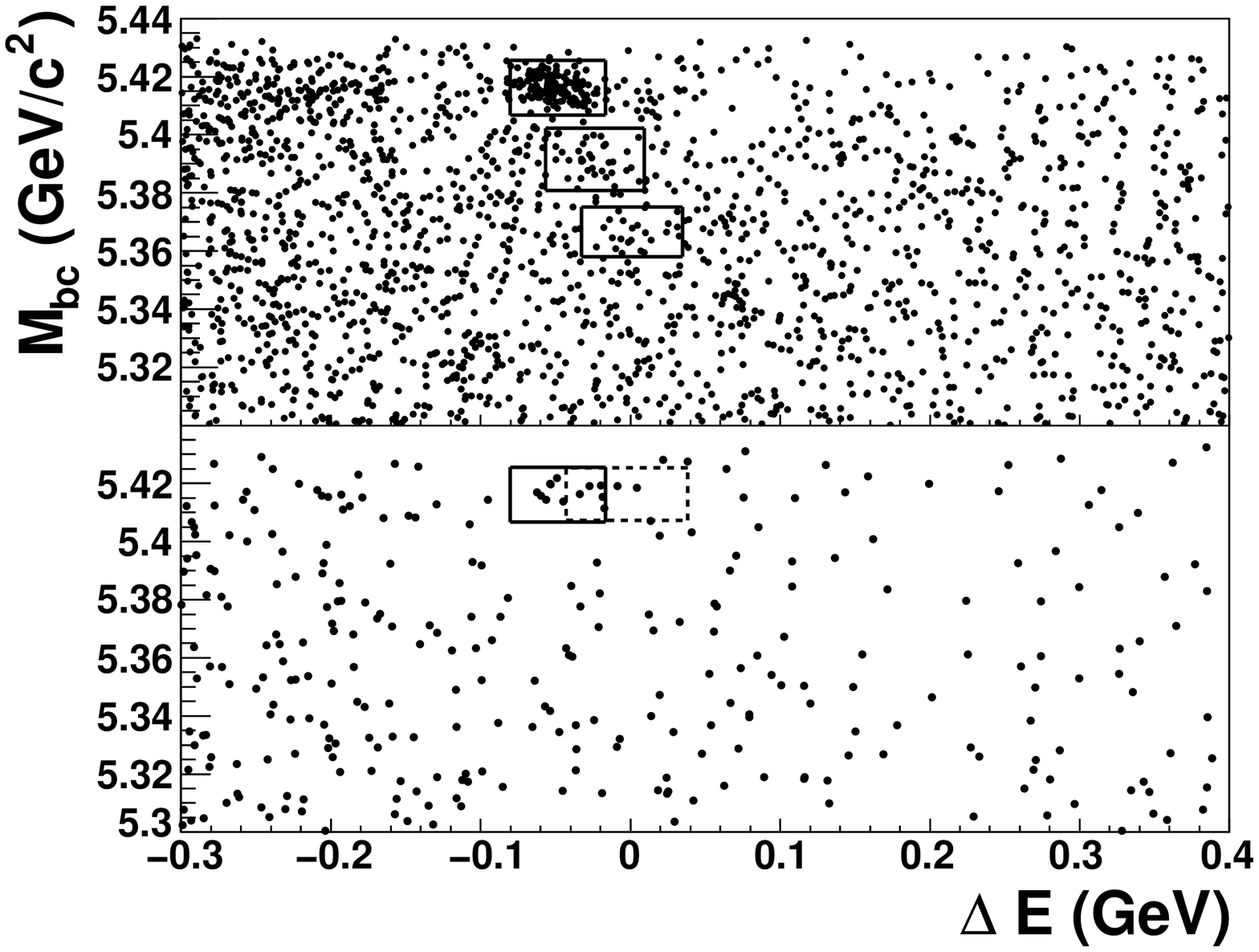}
	\vspace{-0.8cm}
	\caption{$\left(\mbc, \deltae\right)$ scatter plots for $\bsdspi$ (top) and $\bsdsk$ (bottom) candidates.
	  The three boxes in the top plot are the $\pm2.5\,\sigma$ signal regions ($\bsst\barbsst$, $\bsst\bar\bs$ and $\bs\bar\bs$, from top to bottom)
	  while those in the bottom plot are the $\pm2.5\,\sigma$ $\bsst\barbsst$ regions for signal (solid) and for $\bsdspi$ background (dashed).}
	\label{sc:dspi}
      \end{figure}

      To improve signal relative to background, criteria are chosen to maximize  $N_{\textrm{sig}}/\sqrt{N_{\textrm{sig}}+N_{\textrm{bkg}}}$,
      evaluated in the $\bsst\barbsst$ signal region (Fig.~\ref{sc:dspi}).
      Two topological variables are used.
      First, we use the ratio of the second and zeroth Fox-Wolfram moments \cite{PRL_41_1581}, $R_2$,
      which has a broad distribution between zero and one for jet-like continuum events and 
      is concentrated in the range below $0.5$ for the more spherical signal events.
      Candidates for $\bsdspi$ ($\bsdsk$)  are required to have $R_2<0.5$ ($<0.4$).
      We then use the helicity angle $\thetahel$ of the $\ds\to\phi\pi^-$ ($\ds\to K^{\ast0}K^-$) decays,
      defined as the angle between the momentum of the positive daughter of the $\phi$ ($K^{\ast0}$) and
      the momentum of the $\ds$ in the $\phi$ ($K^{\ast0}$) rest frame;
      for signal decays {consisting in a spin--0 particle decaying into a spin--1 particle and a spin--0 particle},
      the distribution is $\propto \cos^2\thetahel$, while for {combinatorial background under $D_s$ signal} it is flat.
      Candidates for $\ds\to\phi\pi^-$ and $\ds\to K^{\ast0}K^-$ are required to
      satisfy $\left|\cos\thetahel\right|>0.2$ ($>0.35$) for the $\bsdspi$ ($\bsdsk$) mode.
      These two selections reject $43\%$ ($73\%$) of the continuum while retaining $95\%$  ($85\%$) of the $\bsdspi$ ($\bsdsk$) signal.
      MC studies show that background from $B^+$ and $B^0$ decays is small and flat enough to be described together with 
      the continuum events for the $\bsdspi$ mode and is negligible for the $\bsdsk$ mode.
      The most relevant background from $\bs$ decays is $\bsdsstpi$.
      
      For each mode, a two-dimensional unbinned extended maximum likelihood fit \cite{NIMA_297_496} in $\mbc$ and $\deltae$ is performed
      on the selected candidates, which are shown in Fig.~\ref{sc:dspi}.
      Each signal probability density function (PDF) is described by a sum of two Gaussians.
      For the $\bsdspi$ analysis, all three $\bs$ production modes ($\bsst\barbsst$, $\bsst\bar\bs$ and $\bs\bar\bs$) are fitted {simultaneously}.
      For the $\bsdsk$ mode, only the $\bsst\barbsst$ component is taken into account.
      The resolutions for $\mbc$ and $\deltae$ are estimated from the MC and scaled by
      a common factor (one for each variable) left free in the $\bsdspi$ fit.
      Approximating $p_{\bsst}^{\ast}$ with $p_{\bs}^{\ast}$ in the $\bsst\to\bs\gamma$ decay,
      the mean values are parameterized, as shown in Table~\ref{param}, as functions of the $\bs$ and $\bsst$ masses,
      which are also left free in the $\bsdspi$ fit.
      The continuum (together with possible $B^+$ and $B^0$ background) is modeled with an ARGUS function \cite{PLB_185_218} for $\mbc$ and
      a linear function for $\deltae$.
      A non-parametric two-dimensional PDF, obtained from MC with the KEYS method \cite{keys}, is used to describe the shape of the $\bsdsstpi$ background.
      
      \begin{table}[ht]
	\caption{Parameterization of $\mbc$ and $\deltae$ mean values.}
	\label{param}
	\begin{ruledtabular}
	  \renewcommand{\arraystretch}{2.5}   
	  \begin{tabular}
            {@{\hspace{0.1cm}}l@{\hspace{0.5cm}}c@{\hspace{0.1cm}}}
            \vspace{-1.1cm}      \\
            Signal          & Mean of $\left(\mbc,\deltae\right)$                                                                          \\
            \hline
            \vspace{.2cm}
            $\bsst\barbsst$    & $\left(m_{\bsst}\,,~\sqrt{\ebeam^2-\left(m_{\bsst}^2-m_{\bs}^2\right)}-\ebeam\right)$                                 \\
            \vspace{-0.25cm}
            $\bsst\bar\bs$   & $\left(\sqrt{\frac{m_{\bsst}^2+m_{\bs}^2}2-\left(\frac{m_{\bsst}^2-m_{\bs}^2}{4\ebeam}\right)^2}\,,~-\frac{m_{\bsst}^2-m_{\bs}^2}{4\ebeam}\right)$ \\
            $\bs\bar\bs$        & $\left(m_{\bs}\,,~0\right)$                                                                                           \\
	  \end{tabular}
	\end{ruledtabular}
      \end{table}
      
      For the $\bsdspi$ mode, the three signal yields are expressed as a function of three free parameters, $\BR\left(\bsdspi\right)$,
      $f_{\bsst\barbsst}$, and $f_{\bsst\bar\bs}$,
      with the relations $N_{\textrm{M}}=N_{\bs}\BR\left(\bsdspi\right)f_{\textrm{M}}\sum_k\varepsilon_k^{\textrm{M}}\BR_k$
      where $\textrm{M}$ is one of the three $\bsSTbsST$-pair production modes and $k$ runs over the $\ds$ modes;
      the third fraction is defined as $f_{\bs\bar\bs}=1-f_{\bsst\barbsst}-f_{\bsst\bar\bs}$.
      The values of $\sum_k\varepsilon_k^{\textrm{M}}\BR_k$,
      which are the total $\ds$ branching fractions \cite{pdg} weighted by the reconstruction efficiencies,
      are listed in Table \ref{eff}.
      
\begin{figure}[t]
	\centering
	\includegraphics[width=0.492\linewidth]{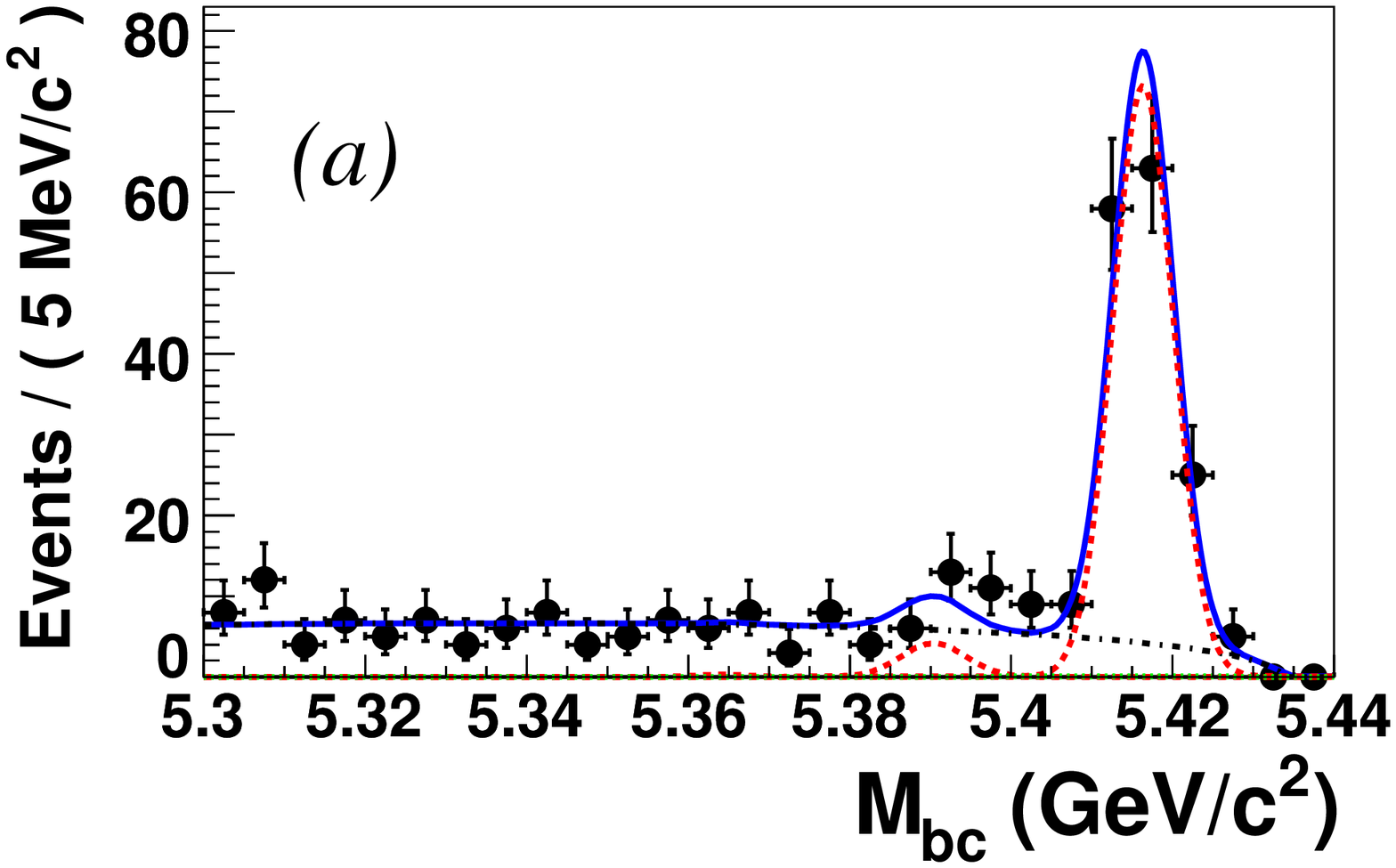}
	\includegraphics[width=0.492\linewidth]{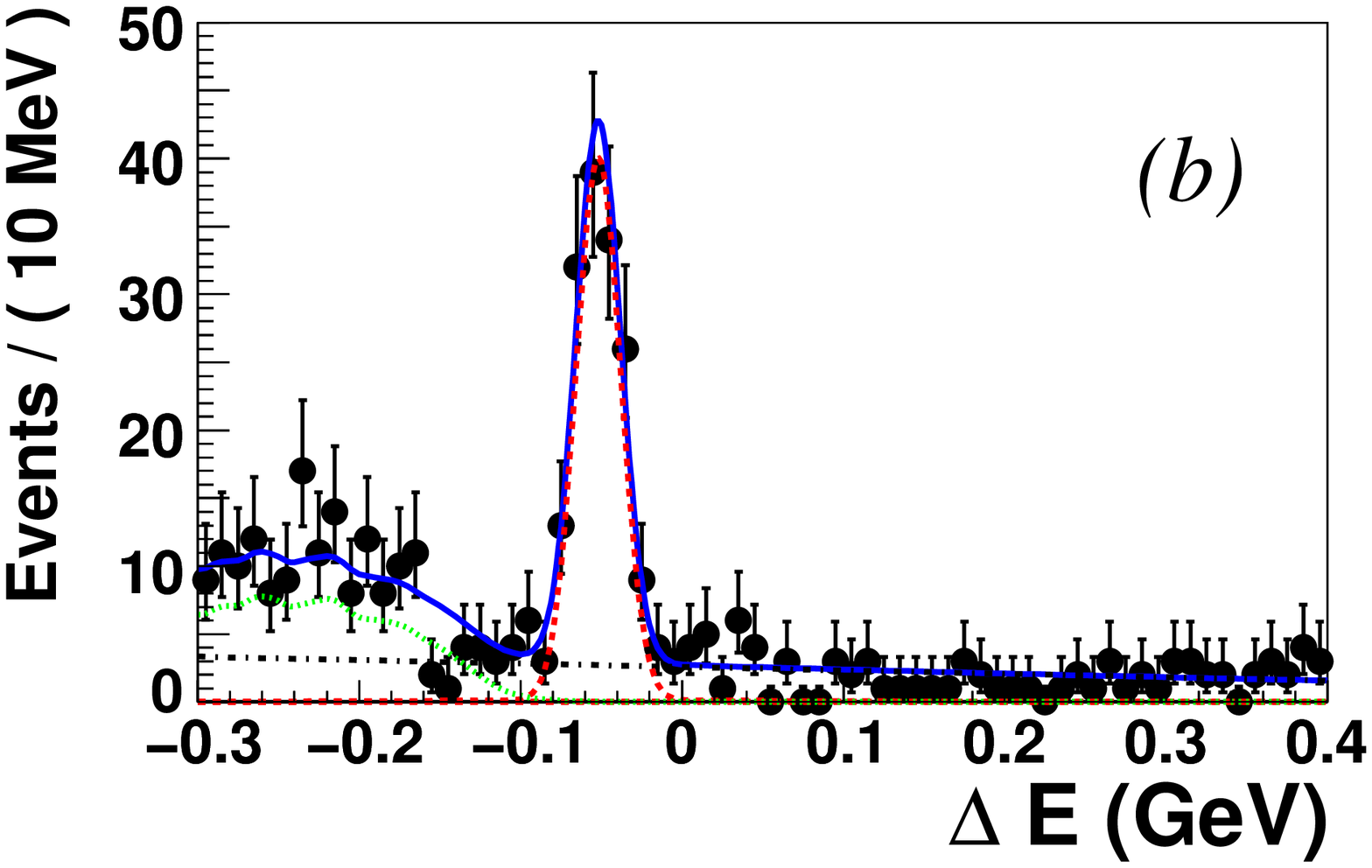}
	
	\vspace{-0.1cm}
	
	\includegraphics[width=0.492\linewidth]{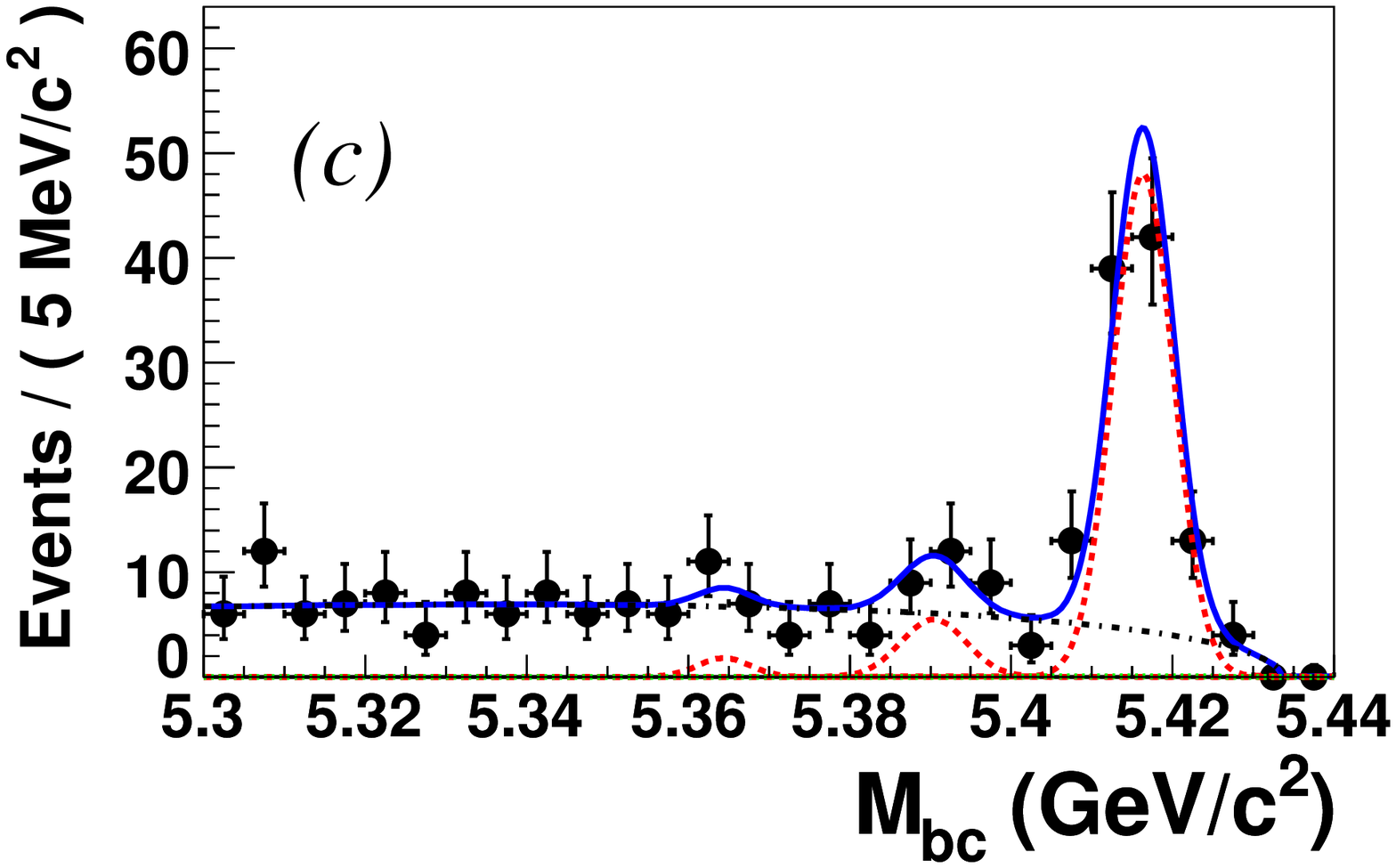}
	\includegraphics[width=0.492\linewidth]{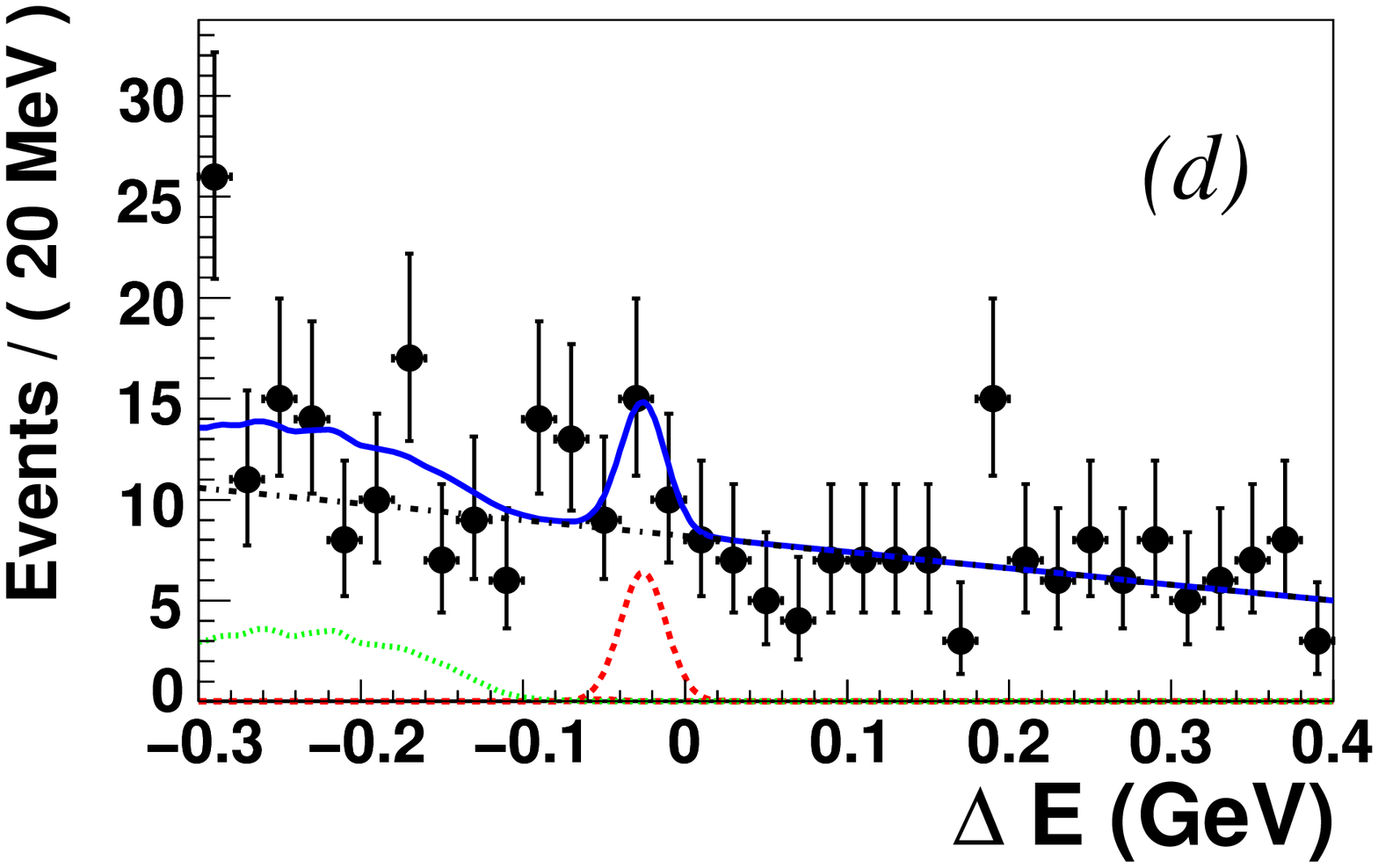}
	
	\vspace{-0.35cm}
	
	\caption{($a$) $\mbc$ distribution of the $\bsdspi$ candidates with
          $\deltae$ in the $\bsst\barbsst$ signal region $[-80,-17]\mev$.
	  ($b$) $\deltae$ distribution of the $\bsdspi$ candidates with
	  $\mbc$ in the $\bsst\barbsst$ signal region $[5.41,5.43]\mgev$. The different fitted components are shown with dashed curves for the signal,
	  dotted curves for the $\bsdsstpi$ background, and dash-dotted curves for the continuum.
	  ($c$) and ($d$) shows the same distributions but using the $\bsst\bar\bs$ signal region ($\deltae\in[-57,9]\mev$ and $\mbc\in[5.38,5.40]\mgev$).}
	\label{dspi:bsstbsstproj}
      \end{figure}

      Figure \ref{dspi:bsstbsstproj} shows the $\mbc$ and $\deltae$ projections in the $\bsst\barbsst$ and
      in the $\bsst\bar\bs$ regions of the data, together with the fitted function.
      In the $\mbc$ distribution, the three signal components are present due to overlap of the signal boxes;
      the peak on the right (middle, left) is due to $\bsst\barbsst$ ($\bsst\bar\bs$, $\bs\bar\bs$) production.
      Table \ref{eff} presents the fitted signal yields as well as the significance defined by
      $S=\sqrt{2\ln\left(\mathcal L_{\textrm{max}}/\mathcal L_0\right)}$ where
      $\mathcal L_{\textrm{max}}$ ($\mathcal L_0$) is the value at the maximum (with the corresponding yield set to zero)
      of the likelihood function convolved with a Gaussian distribution that represents the systematic errors.
      
      \begin{table}[t]
	\caption{Signal efficiencies, yields ($N$) and significances ($S$).}
	\label{eff}
	\begin{ruledtabular}
	  \renewcommand{\arraystretch}{1.3}
	  \begin{tabular}
	    {@{\hspace{0.5cm}}l@{\hspace{0.5cm}}c@{\hspace{0.5cm}}@{\hspace{0.5cm}}c@{\hspace{0.5cm}}@{\hspace{0.5cm}}c@{\hspace{0.5cm}}}
	    $\Upsilon(5S)$ mode  & $\sum_k\varepsilon_k\BR_k$    & $N$                  &  $S$              \\
	    \hline
	    \multicolumn{2}{l}{\hspace{1cm}$\bsdspi$ mode}        & $161\pm15$           &                   \\
	    $\bsst\barbsst$         & 1.58\%                        & $\yieldbsstbssttodspi$ & $\dspisignbsstbsst$ \\
	    $\bsst\bar\bs$           & 1.58\%                        & $\yieldbsstbstodspi$   & $\dspisignbsstbs$   \\
	    $\bs\bar\bs$             & 1.56\%                        & $\yieldbsbstodspi$     & $\dspisignbsbs$    \\
	    \hline
	    \multicolumn{4}{l}{\hspace{1cm}$\bsdsk$ mode}                                                   \\
	    $\bsst\barbsst$         & 1.12\%                        & $\yieldbstodsk$        & $\dsksign$          \\
	  \end{tabular}
	\end{ruledtabular}
      \end{table}
      
      Systematic uncertainties on the branching fractions are shown in Table \ref{sys1}.
      Those on $f_{\bsst\barbsst}$ and $f_{\bsst\bar\bs}$ are mainly due to PDF uncertainties.
      Those due to the beam energy, the momentum calibration and the $p_{\bsst}^{\ast}\approx p_{\bs}^{\ast}$ approximation
      are propagated as systematics on the $\bsst$ mass and $\bs$ mass.
      The momentum normalization uncertainties are much more important in the latter case because
      the measured energy of the $\bs$ candidate is used instead of the beam energy.
      
      \begin{table}[t]
	\caption{Relative systematic uncertainties (in \%) for $\BR\left(\bsdspi\right)$ and $\BR\left(\bsdsk\right)$.}
	\label{sys1}
	\begin{ruledtabular}
	  \renewcommand{\arraystretch}{1.3}
	  \begin{tabular}
	    {@{\hspace{0.1cm}}l@{\hspace{0.5cm}}r@{~~}r@{}@{\hspace{0.5cm}}r@{~~}r@{\hspace{0.2cm}}}
	    Source                            & \multicolumn{2}{c}{$\bsdspi$}      & \multicolumn{2}{c}{$\bsdsk$} \\
	    \hline
	    \hline
	    Integrated luminosity             & $+1.3$    & $-1.3$                       & ${+1.4}$  & ${-1.2}$              \\
	    $\sigma_{b\bar b}^{\Upsilon(5S)}$ & ${+4.8}$  & ${-4.4}$                     & ${+5.0}$  & ${-4.4}$              \\
	    $f_s$                             &  $+13.3$  & $-13.3$                      & $+13.6$   & $-13.4$               \\
	    $f_{\bsst\barbsst}$                  & \multicolumn{2}{c}{\hspace{-3mm}---}     & ${+4.8}$  & ${-4.1}$              \\
	    $\ds$ branching fractions         & ${+6.6}$  & ${-6.1}$                     & ${+6.8}$  & ${-5.9}$              \\
	    Efficiencies (MC stat.)           & $+1.2$    & $-1.2$                       & $+1.5$    & $-1.3$                \\
	    Efficiencies ($R_2$, $\cos\thetahel$)   & $+4.8$    & $-4.8$                       & $+4.8$    & $-4.8$                \\
	    $\pi^{\pm}$, $K^{\pm}$ identification         & $+5.4$    & $-5.4$                       & $+5.2$    & $-5.2$                \\
	    Track reconstruction              & $+4.0$    & $-4.0$                       & $+4.0$    & $-4.0$                \\
	    PDF shapes                        & $+1.0$    & $-1.0$                       & ${+3.3}$  & ${-2.7}$              \\
	    \hline
	    Total                             & $+17.8$   & $-17.5$                      & ${+19.0}$ & ${-18.1}$             \\
	  \end{tabular}
	\end{ruledtabular}
      \end{table}
      
      We measure the branching fraction
      $\BR\left(\bsdspi\right)=\bfbstodspi$ where the largest systematic uncertainty, due to $f_s$, is quoted separately,
      the fraction $f_{\bsst\barbsst}=\fss$
      and the two fitted masses $m_{\bs}=$ $\mbs$ and $m_{\bsst}=$ $\mbsst$.
      These four measurements supersede the previous Belle values \cite{PRD_76_012002}.
      We obtain for the first time values for the two fractions $f_{\bsst\bar\bs}=$ $\fs$ and $f_{\bs\bar\bs}=$ $\f$,
      using the correlation ($-0.77$) between $f_{\bsst\barbsst}$ and $f_{\bsst\bar\bs}$.
      
      Our branching fraction is compatible with the CDF result \cite{pdg,PRL_98_061802},
      {and is slightly higher (1.3$\sigma$) than $\BR(B^0\to D^-\pi^+)$ \cite{pdg}.}
      The value of $f_{\bsst\barbsst}$ is significantly larger than the theoretical expectation of $\approx 70$\% \cite{PRD_55_272,PRL_53_878}.
      The $\bs$ mass is compatible with the world average value \cite{pdg} while our value for the $\bsst$ mass is
      {$2.6\,\sigma$} larger than the result from CLEO \cite{PRL_96_152001}.
      The mass difference obtained, $m_{\bsst}-m_{\bs}=\deltambsstmbs$, is {$4.0\,\sigma$} larger than
      the world average of $m_{B^{\ast0}}-m_{B^0}$ \cite{pdg}{, while heavy-quark symmetry predicts equal values \cite{PRD_68_054024}}.
      
\begin{figure}[ht]
	\centering
	\includegraphics[width=0.492\linewidth]{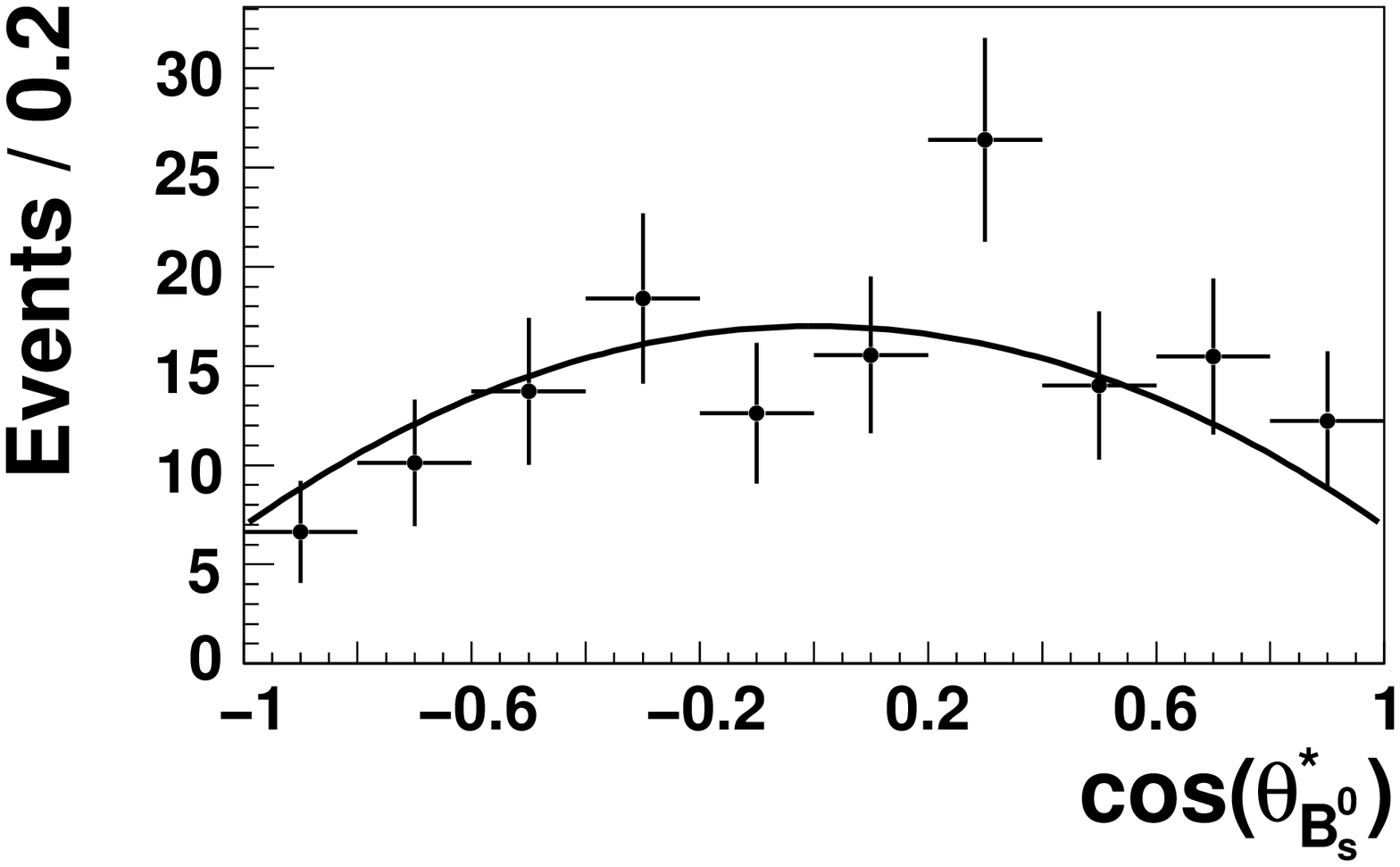}

	\vspace{-0.40cm}
	
	\caption{Fitted distribution of the cosine of the angle between the $\bs$ momentum and
	  the beam axis in the CM frame for the $\Upsilon(5S)\to\bsst\barbsst$ signal.}
	\label{splot}
      \end{figure}

      The distribution of the angle between the $\bs$ momentum and the beam axis in the CM frame is of theoretical interest \cite{PRD_55_272} and
      is presented in Fig.~\ref{splot} for the signal events in the $\bsst\barbsst$ region, using the $_sPlot$ method \cite{NIMA_555_356}.
      A fit to a $1+a\cos^2\theta_{\bs}^{\ast}$ distribution returns $\chi^2/{\textrm{n.d.f.}}=8.74/8$ and $a=-0.59^{+0.18}_{-0.16}$.
      It has been checked that the signal efficiency does not depend on this angle.
      {We naively expect $a=-0.27$ by summing over all the possible polarization states.}
      
      For the $\bsdsk$ mode, mean values and resolutions for $\bsdsk$ and $\bsdspi$ components are calibrated using the results of the $\bsdspi$ fit.
      The four yields (signal, continuum, $\bsdspi$ and $\bsdsstpi$)  are allowed to float, but, due to the very small contribution of $\bsdsstpi$,
      the ratio between the yields of $\bsdsstpi$ and $\bsdspi$ is fixed from a fit to data without kaon identification.
      
\begin{figure}[ht]
	\centering
	\includegraphics[width=0.492\linewidth]{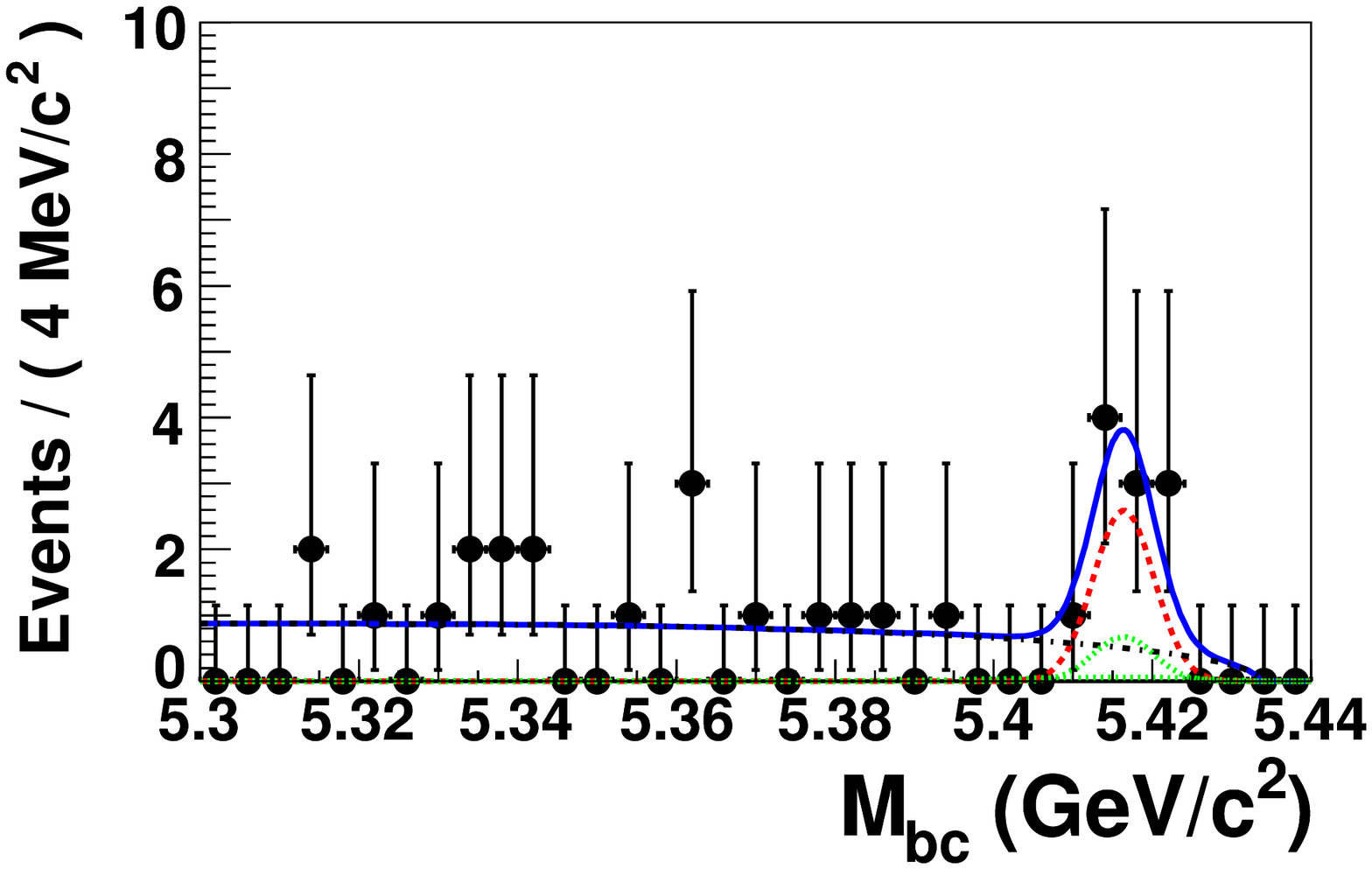}
	\includegraphics[width=0.492\linewidth]{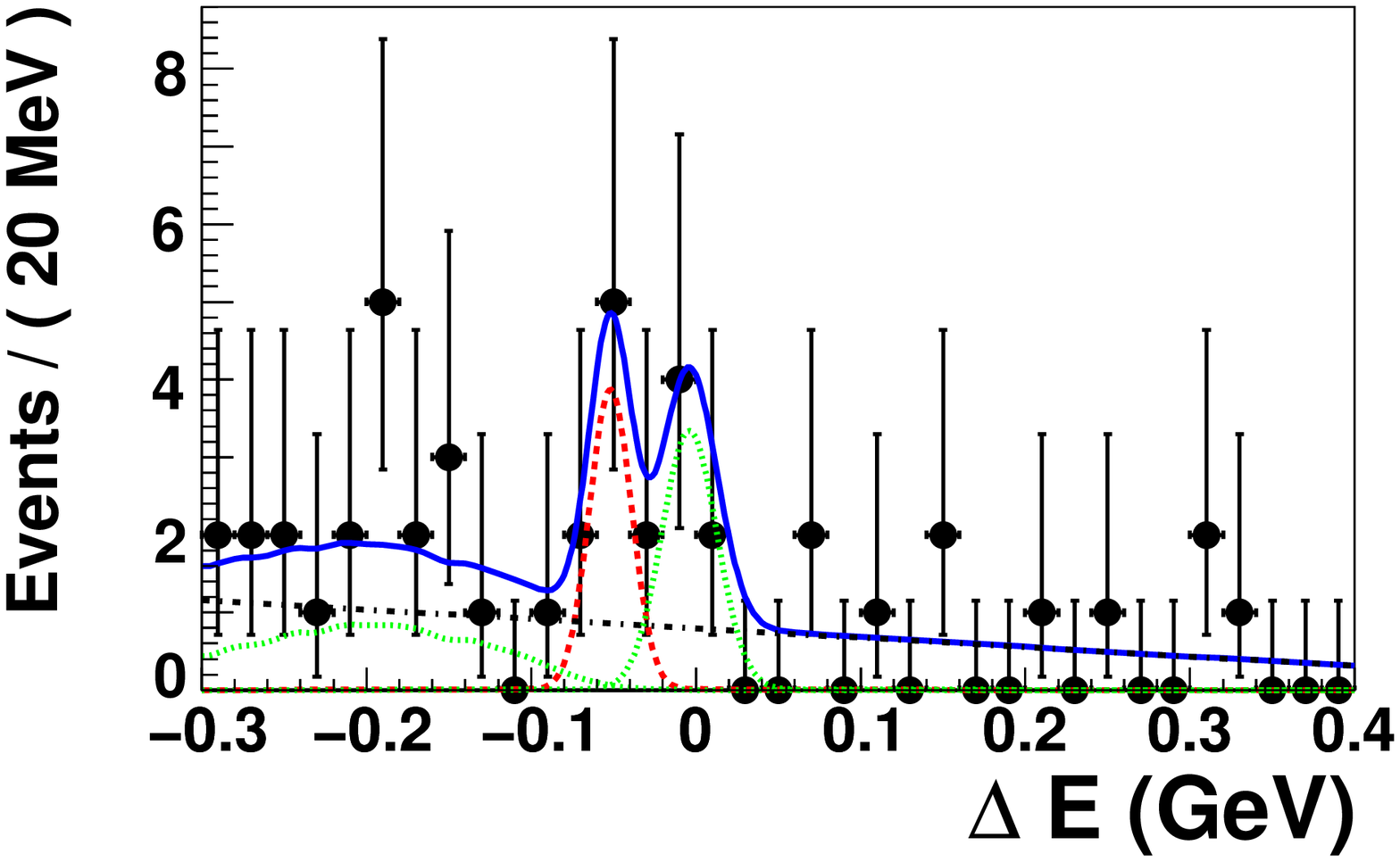}
	
	\vspace{-0.35cm}
	
	\caption{Left: $\mbc$ distribution of $\bsdsk$ candidates with $\deltae$ in the $\bsst\barbsst$ signal region.
	  Right: $\deltae$ distribution of the $\bsdsk$ candidates with $\mbc$ in the $\bsst\barbsst$ signal region;
	  the left (right) peak is the $\bsdsk$ ($\bsdspi$) component.
	  The dashed, dotted and dash-dotted curves represent the signal, $\bsdsSTpi$ backgrounds, and continuum, respectively.}
	\label{dsk:signaldsk}
      \end{figure}

      The fit results are shown in Fig.~\ref{dsk:signaldsk} and Table~\ref{eff}.
      Systematic errors are presented in Table~\ref{sys1}.
      We find $\yieldbstodsk$ signal events ($\dsksign$), corresponding to $\BR\left(\bsdsk\right)$ $=\bfbstodsk$,
      using the previously fitted value of $f_{\bsst\barbsst}$.
      In the ratio $\BR\left(\bsdsk\right)/\BR\left(\bsdspi\right)=\ratio$, the errors are dominated by the low $\bsdsk$ statistics.
      
      In summary, a large $\bsdspi$ signal is observed and six physics parameters are measured: the branching fraction
      $\BR\left(\bsdspi\right)=\bfbstodspiun$ $\bfbstodspideux$, the fractions of the $\bs$ pair production modes at the $\Upsilon(5S)$ energy,
      $f_{\bsst\barbsst}$ $=$ $\fss$,
      $f_{\bsst\bar\bs}=\fs$, $f_{\bs\bar\bs}=\f$, and the masses $m_{\bsst}=\mbsst$, $m_{\bs}=$ $\mbs$.
      In addition, evidence ($\dsksign$) for the $\bsdsk$ decay is obtained,
      leading to a measurement $\BR\left(\bsdsk\right)=$ $\bfbstodskun$ $\bfbstodskdeux$.
      
      \begin{acknowledgments}
	We thank the KEKB group for excellent operation of the
	accelerator, the KEK cryogenics group for efficient solenoid
	operations, and the KEK computer group and
	the NII for valuable computing and SINET3 network
	support.  We acknowledge support from MEXT and JSPS (Japan);
	ARC and DEST (Australia); NSFC (China);
	DST (India); MOEHRD, KOSEF and KRF (Korea);
	KBN (Poland); MES and RFAAE (Russia); ARRS (Slovenia); SNSF (Switzerland);
	NSC and MOE (Taiwan); and DOE (USA).
      \end{acknowledgments}


\begin{thebibliography}{25}
\bibitem[{cha()}]{chargeconj}
	\bibinfo{note}{{Unless specified otherwise, charge-conjugated modes are implied
            throughout the Letter.}}
\end{thebibliography}
\end{document}